\documentclass[%
 aip,
 amsmath,amssymb,
 reprint,%
]{revtex4-1}

\usepackage{indentfirst}
\usepackage{graphicx}
\usepackage{dcolumn}
\usepackage{bm}

\usepackage[utf8]{inputenc}
\usepackage[T1]{fontenc}
\usepackage{mathptmx}
\usepackage{etoolbox}
\usepackage{xcolor}

\makeatletter
\def\@email#1#2{%
 \endgroup
 \patchcmd{\titleblock@produce}
  {\frontmatter@RRAPformat}
  {\frontmatter@RRAPformat{\produce@RRAP{*#1\href{mailto:#2}{#2}}}\frontmatter@RRAPformat}
  {}{}
}%
\makeatother
\begin{document}

\preprint{AIP/123-QED}

\title{Adaptive protocols for SU(11) interferometers to achieve ab initio phase estimation at the Heisenberg limit}

\author{Mingchen Liu}
\affiliation{Department of Physics, Tsinghua University, Beijing 100084, China}

\author{Lijian Zhang}
\affiliation{ National Laboratory of Solid State Microstructures,
Key Laboratory of Intelligent Optical Sensing and Manipulation, College of Engineering and Applied Sciences,
and Collaborative Innovation Center of Advanced Microstructures, Nanjing University, Nanjing 210093, China
}

\author{Haixing Miao}
\affiliation{Department of Physics, Tsinghua University, Beijing 100084, China}

\email{haixing@tsinghua.edu.cn}
\email{lijian.zhang@nju.edu.cn}

\date{\today}
\begin{abstract}
The precision of phase estimation with interferometers can be greatly enhanced using non-classical quantum states, and the SU(11) interferometer is an elegant scheme, which generates two-mode squeezed state internally and also amplifies the signal.   
It has been shown in [Phys. Rev. A {\bf 95}, 063843 (2017)] that the photon-number measurement can achieve the Heisenberg limit, but only for estimating a small phase shift. 
We relax the constraint on the phase size by considering two adaptive protocols: 
one also uses the photon-number measurement with a specially tuned sequence of feedback phase; 
the other implements the yet-to-be-realised optimal measurement but without fine tuning. 
\end{abstract}
\maketitle
\section{Introduction}
Precision measurements of physical quantities are of paramount significance in metrology, imaging, and communication applications, where optical phase estimation serves as a fundamental pillar \,\cite{giovannetti2011advances,ono2013entanglement,simon2017quantum}.
However, as the phase does not correspond to a proper quantum observable, we rely on measuring a phase-dependent quantity, i.e. an estimator , to retrieve it. 
It is highly desirable to develop a scheme that maximizes the precision of phase estimation with a fixed amount of resources. 
Investigating the ultimate limit of phase estimation can be appropriately addressed in the framework of quantum estimation theory \,\cite{helstrom1969quantum,demkowicz2015quantum}. 
It dictates that, for the unbiased estimator, the estimation precision is ultimately bounded by the so-called Quantum $Cram\acute{e}r$-$Rao$ bound (QCRB)\,\cite{braunstein1994statistical,paris2009quantum,yang2019attaining}. 
The QCRB depends on the quantum properties of the probe states, and can be reduced by using non-classical states, e.g. the squeezed states \,\cite{caves1981quantum,anisimov2010quantum,nielsen2021deterministic} and NOON states \,\cite{slussarenko2017unconditional}. 

\begin{figure}[t!]
\includegraphics[width=0.7\columnwidth]{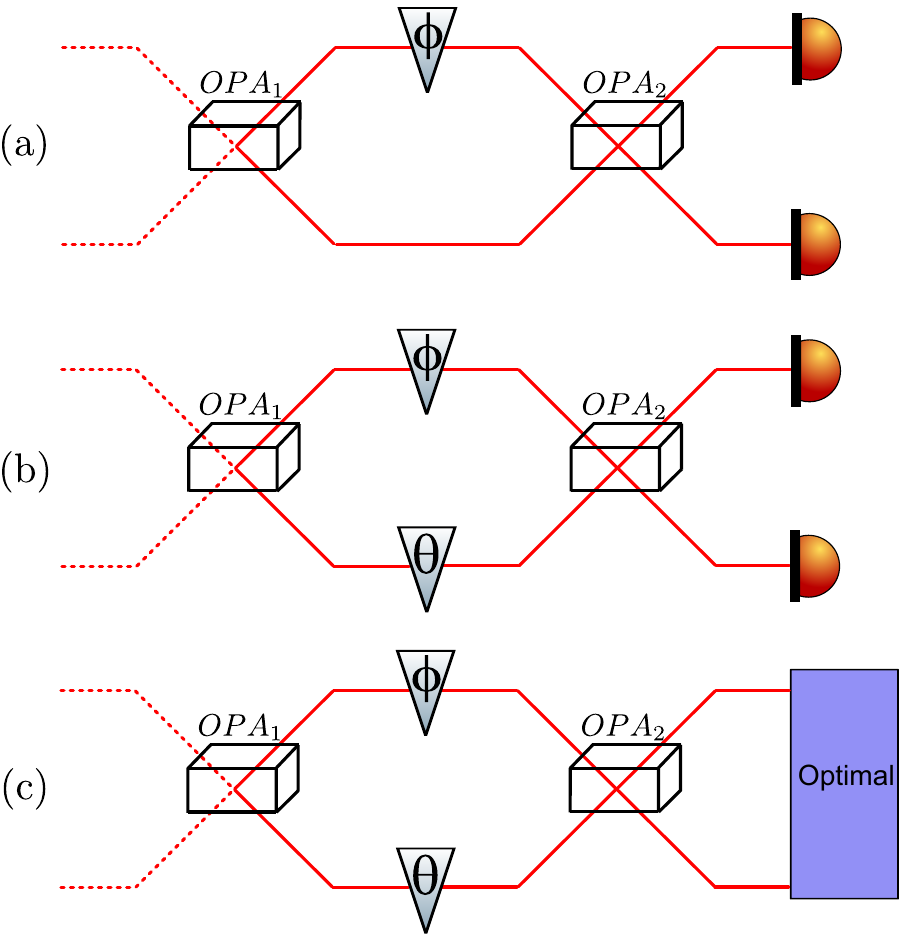}
\caption{The illustration of the vacuum-seeded SU(11) interferometers based on (a) the photon-number measurement, (b) the adaptive photon-number measurement and (c) the adaptive, optimal  measurement.}
\label{fig:SU11}
\end{figure}

\begin{figure}[t!]
\includegraphics[width=8cm]{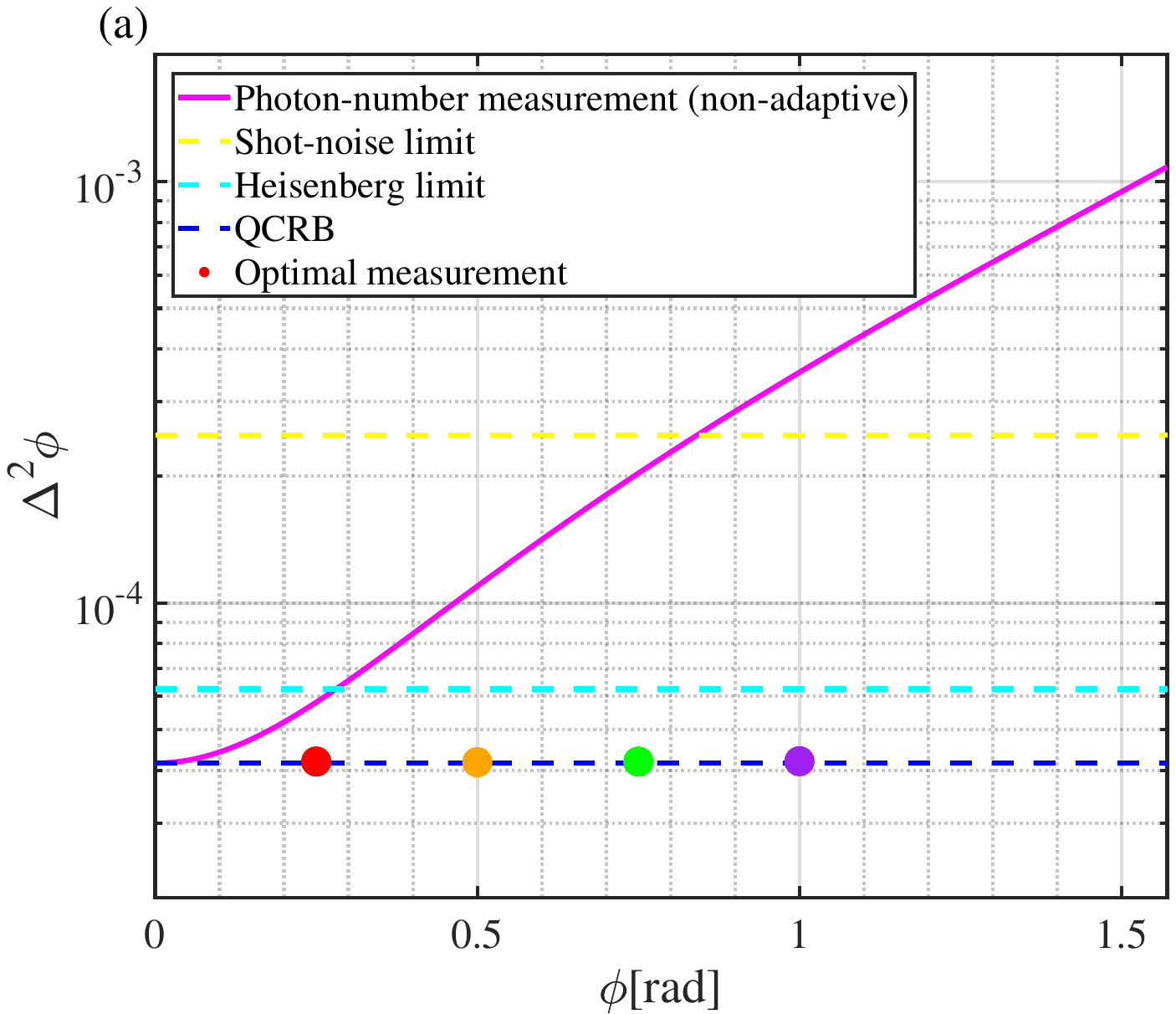}
\includegraphics[width=8cm]{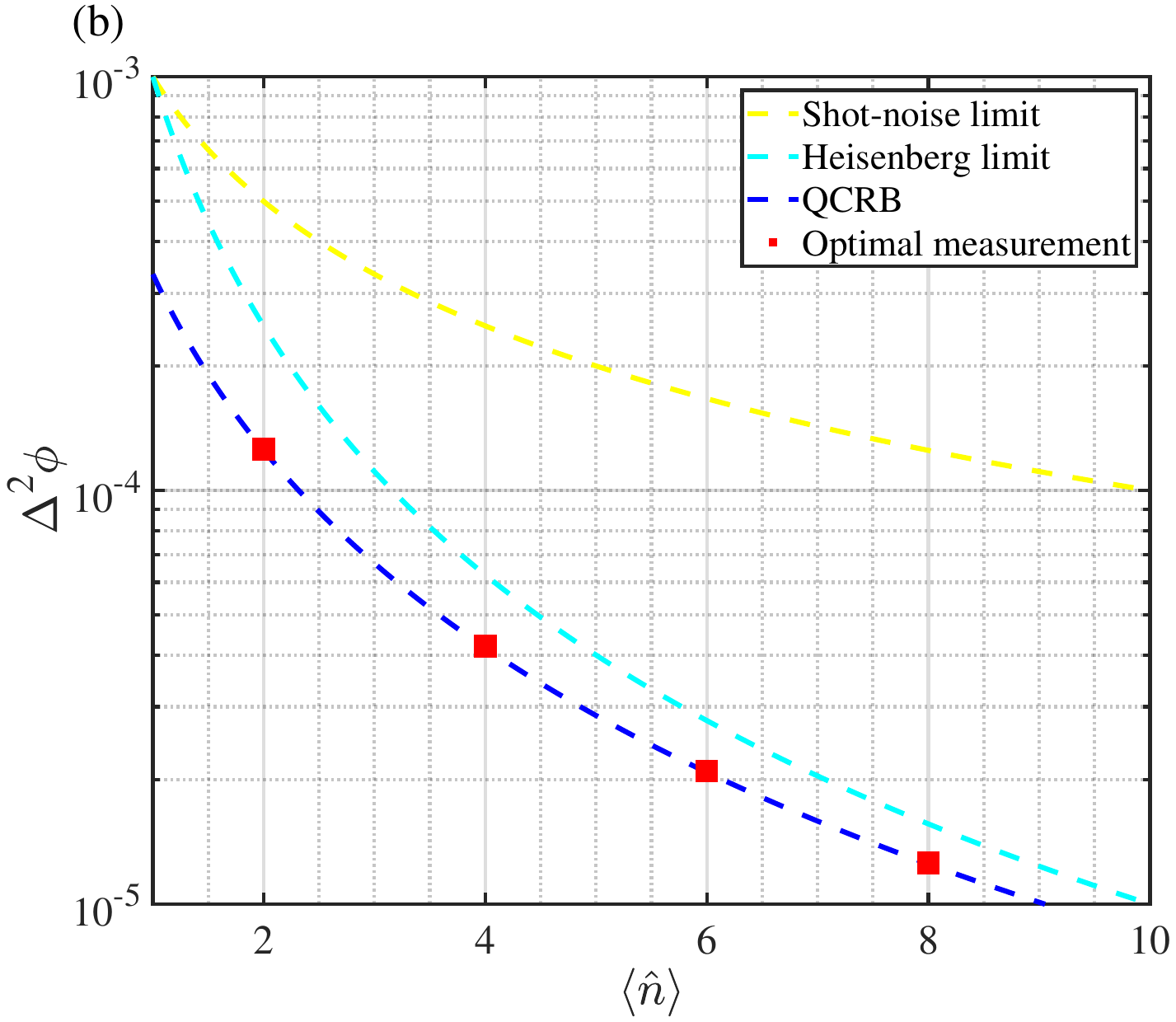}
\caption{The estimation precision of the optimal measurement scheme for $M$=1000 obtained by numerical simulations compared to several precision benchmarks.
(a) The estimation variances of estimating four specific phases based on the optimal measurement scheme for $\langle\hat{n}\rangle$=4, with the red, orange, green and purple scatter points representing the estimation variances of estimating $\phi_t$=0.25, 0.50, 0.75, 1.00 respectively, compared to the shot-noise limit, Heisenberg limit and QCRB.
(b) The estimation variances for four different average photon numbers based on the optimal measurement scheme for $\phi_t$=0.75, with the four red scatter points from left to right representing the estimation variances for $\langle\hat{n}\rangle$=2, 4, 6, 8 respectively, compared with the shot-noise limit, Heisenberg limit and QCRB.
}
\label{fig:D2phi}
\end{figure}

In this paper, we consider the vacuum-seeded SU(11) interferometer, in which the non-classical two-mode squeezed state is generated by the nonlinear parametric process \,\cite{yurke19862,li2014phase,ou2020quantum,du20222}.
As illustrated by Fig.\,\ref{fig:SU11}, the SU(11) interferometer is a Mach-Zehnder interferometer (MZI) with the beam splitter replaced by the nonlinear optical parametric amplifier (OPA).
The standard measurement, shown in Fig.\,\ref{fig:SU11}(a), uses photon detectors to count photons and the phase estimator is based upon the measured photon number in either output port. 
As illustrated  by Fig.\,\ref{fig:D2phi}(a), it turns out that such a measurement is only optimal when the phase difference between the two arms is close to zero, and for ab initio estimation of an arbitrary unknown phase, the precision significantly decreases\,\cite{anderson2017optimal}. 
To recover the precision, we propose using an adaptive measurement scheme shown in Fig.\,\ref{fig:SU11}(b). 
A feedback phase $\theta$ is introduced in one arm, and conditional on the outcomes of the photon-number measurement, it is tuned to minimise the phase difference between the two arms\,\cite{berni2015ab,liu2017quantum,zheng2019ab}.
However, because the likelihood function for the photon-number measurement is even with respect to the phase difference, the posterior probability distribution for the estimator is bimodal with two peak values, which leads to an ambiguity in the estimation. 
We need to design a special, non-unique sequence for tuning the feedback phase at each step. 
To solve this issue at the fundamental level, we derive the optimal measurement operator using quantum estimation theory, without any ambiguity in the likelihood function. 
It is illustrated in Fig.\,\ref{fig:SU11}(c), and the optimal measurement operator is given by\,\cite{gaiba2009squeezed} 
\begin{equation}
\hat L_0=\sqrt{\langle\hat{n}\rangle(\langle\hat{n}\rangle+2)}\left[|0,0 \rangle\langle 1,1|+|1,1 \rangle\langle 0,0|\right]\,, 
\label{eq:L0}
\end{equation}
where $\langle\hat{n}\rangle$ represents the average photon number after the first OPA. 
We can see from Fig.\,\ref{fig:D2phi}(a) that the optimal scheme allows us to reach the QCRB for arbitrary unknown phases\,\cite{you2019conclusive}: 
\begin{equation} 
\Delta^2\phi_{QCRB}=\frac{1}{M\langle\hat{n}\rangle(\langle\hat{n}\rangle+2)}\,.
\label{eq:qcrb}
\end{equation}
Apart from the scaling with respect to the number of measurements $M$, it has the same scaling as the Heisenberg limit with respect to the average photon number, which is demonstrated in Fig.\,\ref{fig:D2phi}(b). 
The question remained is finding the physical realisation of such an optimal estimator, which is an interesting topic for future study. 

We organize the paper as follows. 
In Sec.\,\ref{sec:photon}, we present the adaptive mechanism based upon Bayesian estimation using the photon-number measurement. We show that the likelihood function is even with respect to the phase difference, which leads to the issue of ambiguity due to a bimodal posterior distribution. To solve this issue, we intentionally make the feedback phase smaller than the half of the conditional value for the first few steps. This allows us to break the  symmetry in the bimodal distribution, i.e. the probability for the true phase value and the mirrored one being unequal. 
In Sec.\,\ref{sec:optimal}, we construct the optimal measurement scheme based on quantum estimation theory and demonstrate its performances via numerical simulations.
Sec.\,\ref{sec:conclusion} provides conclusive remarks and explores potential avenues for the physical implementation of the optimal measurement scheme.
\section{Adaptive photon-number measurement}
\label{sec:photon}

The adaptive measurement protocol is based upon the Bayesian inference; the feedback phase is adjusted according to the posterior probability distribution, which is defined as\,\cite{olivares2009bayesian}: 
\begin{equation}
p(\phi|\xi)=\frac{p(\xi|\phi)p(\phi)}{p(\xi)}\,.
\end{equation}
Here $p(\phi)$ is the the prior distribution and $p(\xi|\phi)$ is the likelihood function  with "$\xi$" denoting the measurement outcome.
The value of the feedback phase for each step is determined by the current estimate $\phi_{est}$, which is obtained by selecting the maximum value of the posterior probability distribution from the previous measurement round. 

However, for the photon-number measurement, the likelihood function for detecting $n$ photons at both output ports is an even function with respect to the arm phase difference $\phi-\theta$:
\begin{equation}
p(n|\phi) = \sum^{\infty}_{p=0}\sum^{\infty}_{q=0}
D(0,n;n,0)\cos[(p-q)(\phi-\theta)]\,,
\end{equation}
where $D(0,n;n,0)$ is the coefficient with its detail shown in the Appendix \ref{app:lf}. 
Therefore, if we set the feedback phase $\theta$ in one of the arms, we cannot tell the difference between the two phase values of $\phi$ and $2\theta-\phi$ in the other arm. In another word, the posterior probability distribution is bimodal with two peaks at the true phase $\phi_{t}$ and $2\theta-\phi_t$, respectively. 

\begin{figure}[h!]
\includegraphics[width=8cm]{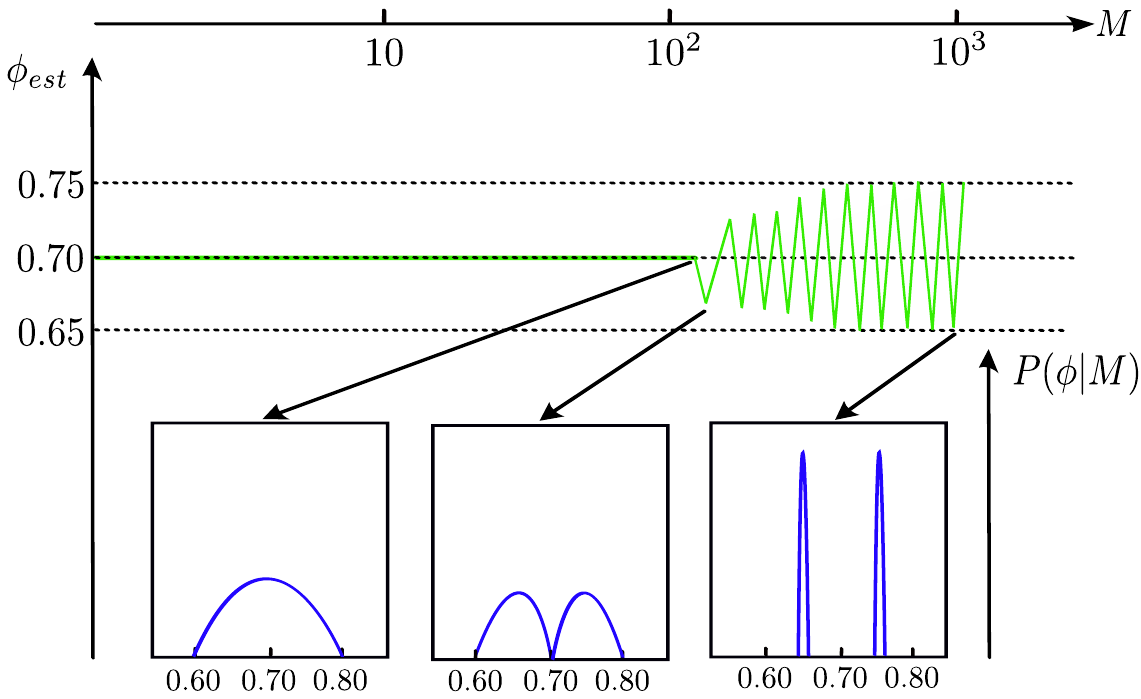}
\caption{The iteration of estimates $\phi_{est}$ with the number of measurements in the case of $\theta=0.70$ and $\phi_t=0.75$ for $\langle\hat{n}\rangle=4$ and the conditional posterior probability distributions for $M$=$M_{\rm threshold}$-1, $M_{\rm threshold}$ and 1000, which are all qualitative diagrams}
\label{fig:estM}
\end{figure}

We demonstrate this ambiguity explicitly by using the
numerical simulation, of which the result is illustrated in Fig.\,\ref{fig:estM}. 
In the simulation, we set $\theta=0.70$ and $\phi_t=0.75$.
The real-time estimate $\phi_{est}$ does not converge to the true phase value as the number of measurements $M$ increases; instead, it displays oscillations beyond a certain threshold value of $M$, denoted as $M_{\rm threshold}$. 
When the estimation error becomes of the order of unity---the separation of two peaks, the posterior probability distribution undergoes a transition from unimodal to bimodal with two peaks shown up at $\phi=0.65$ and $\phi=0.75$, respectively. 
Furthermore, $M_{\rm threshold}$ exhibits an increase as $\theta$ approaches $\phi_t$, as is shown in Table \ref{tab:table1}. 
This is expected because the statistical error needs to be small enough in order to resolve the two peaks in the posterior distribution.

\begin{table}[t!]
\caption{\label{tab:table1}
$M_{\rm threshold}$ for some specific $\theta$ in the case of $\phi_t=0.75$ and $\langle\hat{n}\rangle$=4.}
\begin{ruledtabular}
\begin{tabular}{ccc}
 $\phi_t$ & $\theta$ & $M_{\rm threshold}$ \\
\hline
0.75 & 0.65 & 10-20 \\ \hline
0.75 & 0.70 & 100-150 \\ \hline
0.75 & 0.74 & 700-900 \\ \hline
0.75 & 0.745 & >1000 \\
\end{tabular}
\end{ruledtabular}
\end{table}

\begin{figure}[t!]
\includegraphics[width=8cm]{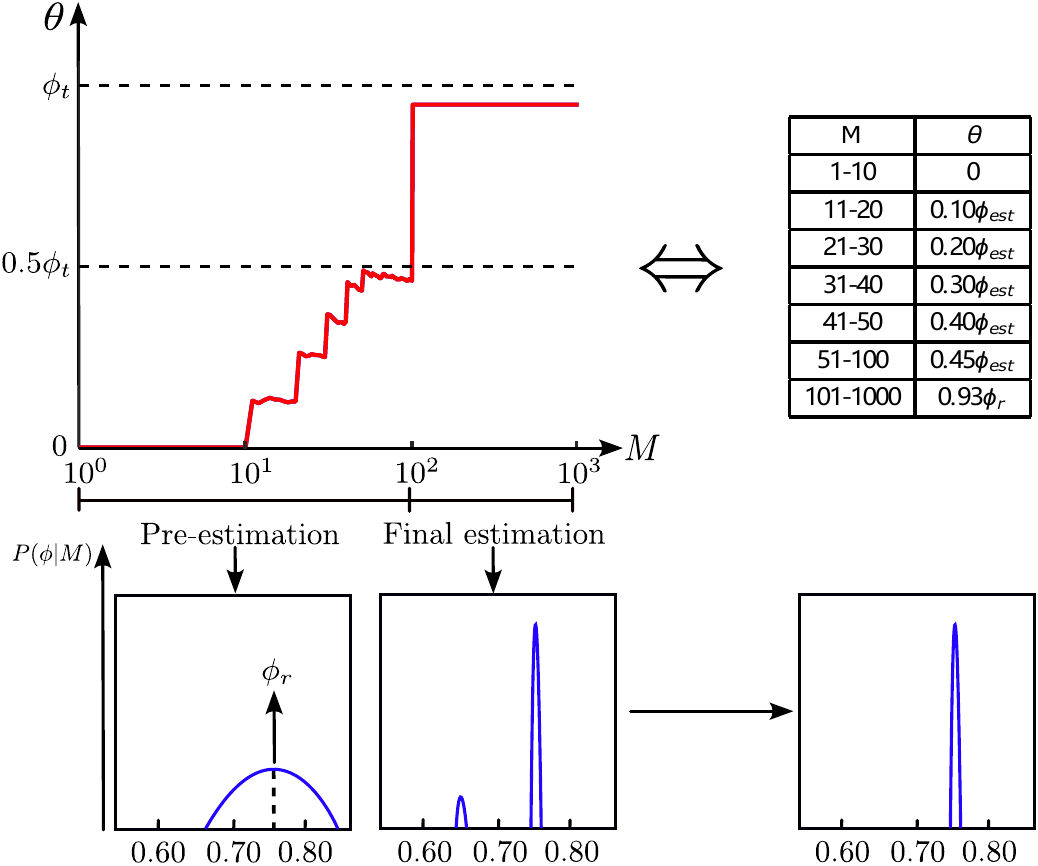}
\caption{The qualitative diagram of how the "ladder" adaptive protocol works. 
The upper panel of the figure depicts how we set the value of the feedback phase $\theta$ for each round of measurements.
The bottom panel of the figure shows the conditional posterior probability distributions given by pre-estimation and final estimation, as well as the final posterior probability distribution obtained after removing the extraneous peak and normalization.}
\label{fig:ladder}
\end{figure}

To overcome the bimodal issue that leads to ambiguity, we propose a modified adaptive protocol, referred to as the "ladder" adaptive protocol, which is illustrated in Fig.\,\ref{fig:ladder}.
For the first stage of measurements, a set of $M_r$ (of the order of $100$) measurements, we make the feedback phase $\theta$ smaller than the one from the maximum likelihood estimate, and gradually increase its value. To guarantee there is no ambiguity at this stage, the value of $\theta$ is constrained to a range that is smaller than $0.5\phi_{est}$. This allows us to create an asymmetry between $\phi_t$ and $2\theta-\phi_t$, and enables pre-estimation to yield a reliable rough estimate $\phi_r$ that is close to $\phi_t$.
The stage of final estimation is performed with $\theta=0.93\phi_r$ to achieve the desired emergence of a bimodal posterior probability distribution within a limited number of measurements, while maintaining the estimation precision that is not significantly worse than the QCRB.
Subsequently, another 900 rounds of measurements are conducted to obtain a bimodal posterior probability distribution with two distinct peaks, with the higher-probability peak located at $\phi_t$. 
In the last step, we remove the lower-probability peak and normalize the distribution.
The resulting estimation variance for the final posterior probability distribution is only slightly larger than the QCRB.

\section{Optimal measurement scheme}
\label{sec:optimal}

In addition to the photon-number measurement, we also construct an optimal measurement scheme, which is based upon the symmetric logarithmic derivative (SLD) operator in quantum estimation theory (refer to Appendix \ref{app:optimal} for a brief introduction). 
The corresponding  measurement operator $\hat L_0$ is shown in Eq.\,\eqref{eq:L0}.
It turns out that the new scheme is immune from the issue of ambiguity, resulting in a straightforward adaptive protocol without the need of fine-tuning the feedback phase. 
Furthermore, the accuracy will asymptotically approach the QCRB: 
\begin{equation}
\Delta^2\phi_{\hat{L}_0}|_{\theta\rightarrow\phi_t}
=
\Delta^2\phi_{QCRB}\left[1+{\cal O}(\phi_t-\theta)^2\right]\,. 
\label{eq:Ophi2}
\end{equation}
The additional error is of the second order of the difference between the feedback phase $\theta$ and the true value $\phi_t$, as shown in Appendix \ref{app:veri}. It is of the same order as the QCRB, and is negligible when the product of the mean photon number and the number of measurements is large. 


\begin{figure}[t!]
\includegraphics[width=0.5\columnwidth]{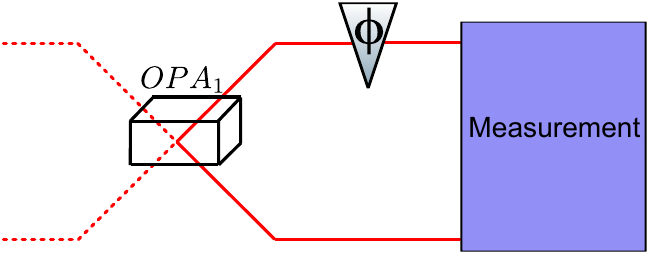}
\caption{The illustration of the truncated vacuum-seeded SU(11) interferometer.}
\label{fig:trun}
\end{figure}

The derivation of the SLD operator involves two starting points: 
the probe state and the generator $\hat G$ associated with the 
phase $\phi$, which is usually defined through the 
following unitary operator: 
\begin{equation}
\hat U(\phi)=e^{-i\phi\hat G}\,. 
\end{equation}
For the SU(11) interferometer, we cannot translate the 
phase generation process into the form shown above and define $\hat G$ properly. Instead, we first consider 
the truncated vacuum-seeded SU(11) interferometer as depicted in Fig.\,\ref{fig:trun}. In this case, the probe state is simply the two-mode squeezed state, and the generator $\hat{G}$ is precisely represented by the photon-number operator $\hat{a}^{\dagger}\hat{a}$. The corresponding SLD operator is given by 
\begin{equation}
\hat{L}_\phi=2i\sum_{jk} 
\hat{U}(\phi)
(G_{00jk}|\varphi_{00}\rangle\langle\varphi_{jk}|-G_{jk00}|\varphi_{jk}\rangle\langle\varphi_{00}|)
\hat{U}^{\dagger}(\phi)\,.
\end{equation}
Here $|\varphi_{jk}\rangle = \hat U_{\rm OPA_1}|j,k\rangle$ and $G_{jkmn}=\langle\varphi_{jk}|\hat{G}|\varphi_{mn}\rangle$. Using the expression for $\hat G$ and the orthogonality of the Fock state, we obtain 
\begin{equation}
\hat{L}_\phi
=
\hat{U}(\phi)
\hat{U}_{\rm OPA_1}\hat L_0\,\hat{U}^{\dagger}_{\rm OPA_1}
\hat{U}^{\dagger}(\phi)\,, 
\end{equation}
which is equal to $\hat L_0$ under the inverse operation of $\hat U(\phi)$ and $\hat U_{\rm OPA_1}$.
It is known from quantum estimation theory that the QCRB is saturated if $\hat L_{\phi}$ is measured when $\phi$ is close to the true value $\phi_t$. 
Therefore, since the second OPA in the SU(11) interferometer is usually in the inverse operation regime of the first OPA, we only need to measure $\hat L_0$ and make the feedback phase close to $\phi_t$; and the latter is automatically achieved by the adaptive protocol using the Bayesian estimation. 

\begin{figure}[t!]
\includegraphics[width=8cm]{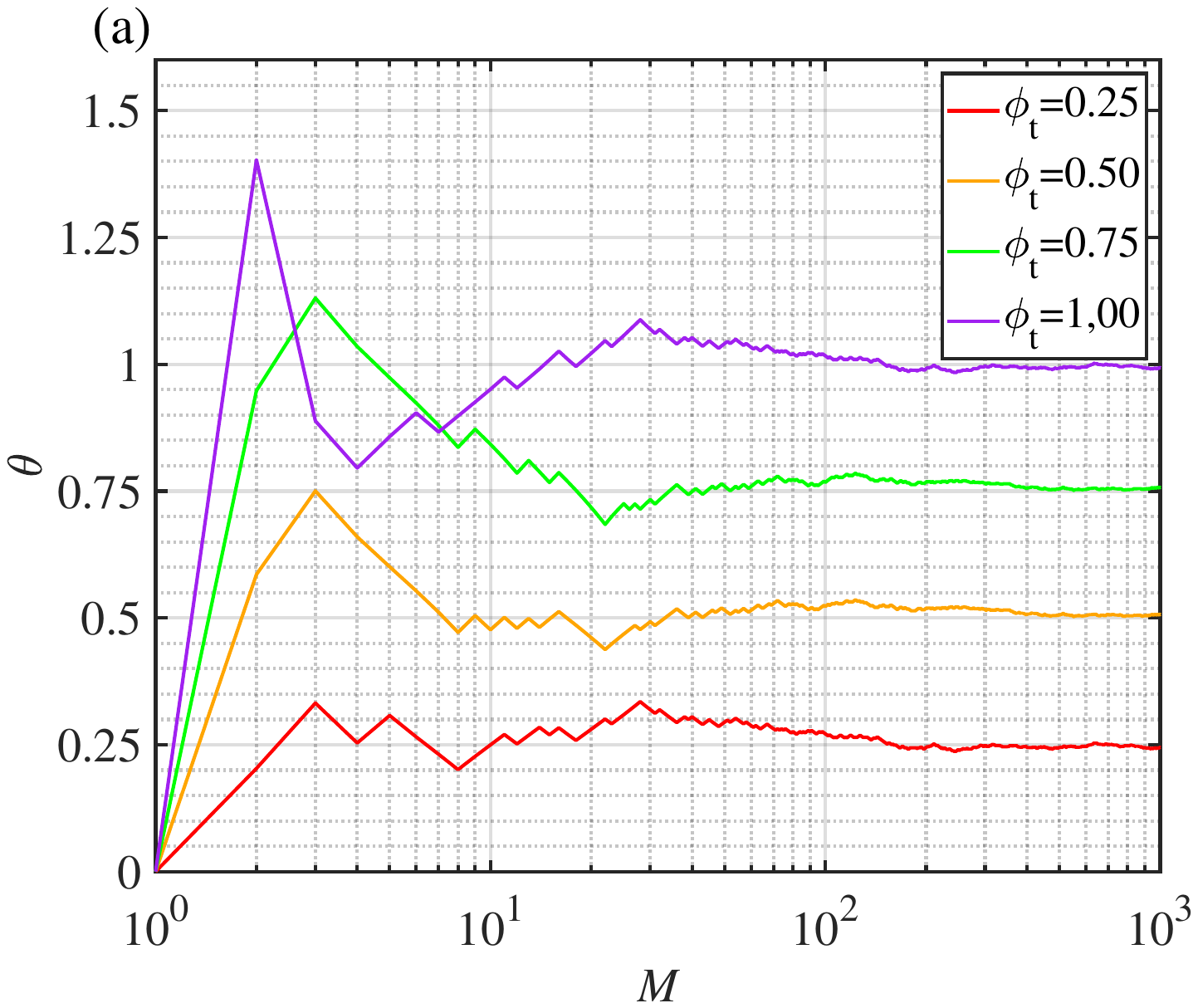}
\includegraphics[width=8cm]{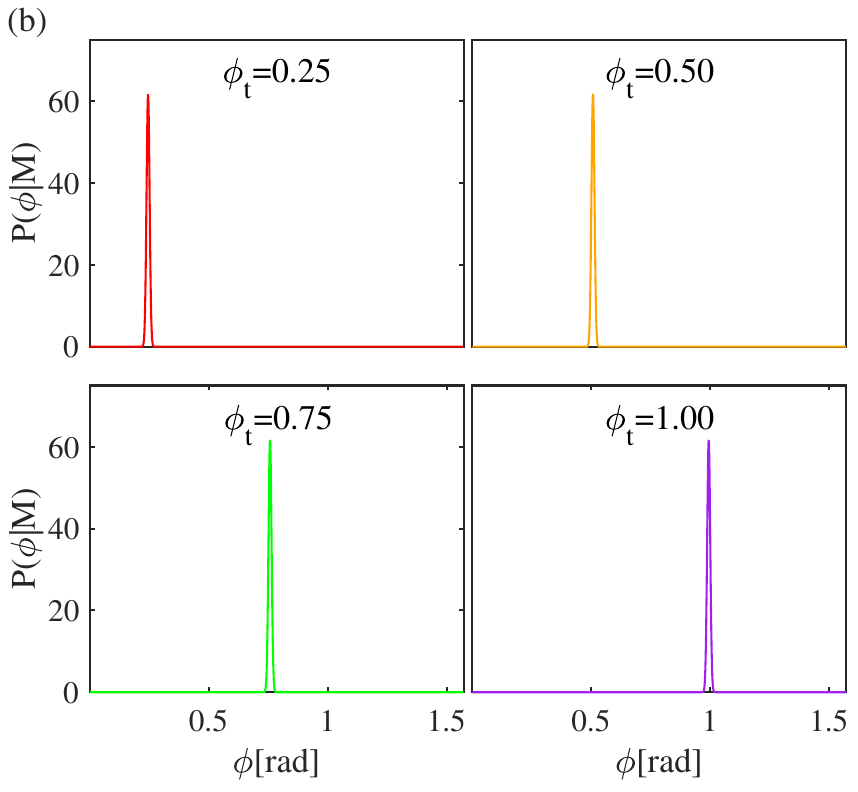}
\caption{The relevant simulation results of estimating $\phi_t=0.25, 0.50, 0.75, 1.00$ for $M=1000$ and $\langle\hat{n}\rangle=4$ based on the optimal measurement scheme. (a) The iteration of $\theta$ with the number of measurements for the four cases. (b) The conditional posterior probability distributions $P(\phi|M)$ that we end up with in the four cases.}
\label{fig:sim}
\end{figure}

For illustration, we perform a similar numerical simulation for the optimal measurement scheme. 
Fig.\,\ref{fig:sim} depicts the relevant simulation results of estimating unknown phase shifts $\phi_t=0.25, 0.50, 0.75, 1.00$ for $M=1000$ and $\langle\hat{n}\rangle=4$.
Specifically, Fig.\,\ref{fig:sim}(a) illustrates the relationships between $\theta$ and the number of measurements for the four cases, respectively showcasing the tendency to converge to the true value of phase.
Fig.\,\ref{fig:sim}(b) displays the conditional posterior probability distribution finally obtained in the four cases as a function of $\phi$. 
The corresponding accuracy for the optimal measurement scheme has been shown in Fig.\,\ref{fig:D2phi}, which indeed follows 
the Heisenberg scaling. 

\section{Conclusion and Outlook}
\label{sec:conclusion}

In this study, we consider two adaptive protocols in the SU(11) interferometer to attain the estimation precision that saturates the QCRB at any phase. 
Both protocols are based upon the Bayesian estimation, in which the posterior distribution defines the feedback phase. 
The first scheme utilizes the photon-number measurement. It requires a specifically designed sequence for fine-tuning the feedback phase to avoid the ambiguity due to the posterior distribution being bimodal. 
The second scheme is built upon the symmetric logarithmic derivative operator in quantum estimation theory; the ambiguity is naturally absent. 
However, the corresponding measurement operator is not straightforward to be implemented because the physical realisation has not yet been found. It remains an open question if we could realise such an optimal measurement scheme by leveraging the advancement in the quantum information and computing theory.  
\begin{acknowledgments}
M. L. and H. M. are supported by State Key Laboratory
of Low Dimensional Quantum Physics and the start-up
fund from Tsinghua University. L. Z. acknowledges the 
support from the National Key Research and Development Program of China (Grant
Nos. 2017YFA0303703 and 2018YFA030602) and the National Natural Science Foundation of
China (Grant Nos. 91836303, 61975077, 61490711 and 11690032) and Fundamental Research
Funds for the Central Universities (Grant No. 020214380068).
\end{acknowledgments}
\appendix

\section{Calculation of the likelihood function}
\label{app:lf}

\subsection{Adaptive photon-number measurement}
The eigenstates of the photon-number operator associated with non-zero measurement probabilities are precisely Twin-Fock states, denoted as $|n,n\rangle$ for $n = 0,1,2,\cdots$. 
Then the probability of observing $n$ photons at both output ports can be expressed as
\begin{equation}
p(n|\phi)=
\langle 0,0|\hat{U}^{\dagger}|n,n\rangle\langle n,n|\hat{U}|0,0\rangle\,,
\label{eq:p_n}
\end{equation}
in which $\hat{U}=\hat{U}_{\rm OPA_2}\hat{U}(\delta\phi)\hat{U}_{\rm OPA_1}$ with $\delta\phi$ being a quantity defined as $\phi-\theta$.
Moreover, $\hat{U}_{\rm OPA}|n,n\rangle$, the so-called two-mode squeezed state, can be expanded as a superposition of Twin-Fock states: $\hat{U}_{\rm OPA}|n,n\rangle=\sum_{m=0}^{\infty}C_m(n)|m,m\rangle$, and the Schmidt coefficient $C_m(n)$ is given by\,\cite{hou2020optical}
\begin{equation}
\begin{aligned}
C_m(n)
& =\sum_{k=0}^{min[m,n]}
\frac{(-1)^{n-k}m!n!sinh^{-2k}r}{(k!)^2(m-k)!(n-k)!}
\frac{tanh^{m+n}r}{coshr}
e^{i(m-n)(\psi-\pi)}\\
& =
C_m^{'}(n)
e^{i(m-n)(\psi-\pi)}\,,
\end{aligned}
\end{equation}
in which $r$ and $\psi$ describe the strength and phase shift in the $\rm OPA$ process.
By applying the above expansion, Eq.\,\eqref{eq:p_n} becomes
\begin{equation}
p(n|\phi)=
\sum_{p=0}^{\infty}\sum_{q=0}^{\infty}
D(0,n;n,0)cos[(p-q)\delta\phi]\,,
\end{equation}
where $D(a,b;c,d)$ is a coefficient defined as $C_p^{'}(a)$ $C_{p}^{'}(b)C_q^{'}(c)C_{q}^{'}(d)$.

\subsection{Optimal measurement}
Let $|\xi_n\rangle$ be an eigenstate of $\hat{L}_{\phi}$, which can always be expressed as $|\xi_n\rangle=\hat{U}|\xi^{'}_n\rangle$.
Then $|\xi^{'}_n\rangle$ is an eigenstate of $\hat{L}_0$.
In the Twin-Fock state representation, it is evident that $\hat{L}_0$ possesses exclusively two eigenstates with non-zero eigenvalues, which are given by 
$|\xi^{'}_{\pm 1}\rangle=\frac{1}{\sqrt{2}}(|0,0 \rangle \pm |1,1 \rangle)$.
Additionally, the other eigenstates whose eigenvalues are zero are $|\xi^{'}_{n}\rangle=|n,n \rangle$ $(n=2,3,\cdots)$.
Finally, the corresponding likelihood functions are given by
\begin{equation}
\begin{aligned}
p(\xi_{\pm 1}|\phi)=
\frac{1}{2}
\sum_{p=0}^{\infty}
\sum_{q=0}^{\infty}
& \{
D(0,0;0,0) \cos[(p-q)\delta\phi]\\
&+
D(0,1;1,0)\cos[(p-q)\delta\phi]\\
&\pm
2D(0,0;1,0)\sin[(p-q)\delta\phi]
\}\,,
\end{aligned}
\end{equation}
\begin{equation}
p(\xi_n|\phi)=\sum_{p=0}^{\infty}\sum_{q=0}^{\infty}
D(0,n;n,0)\cos[(p-q)\delta\phi]
\quad
(n=2,3,\cdots)\,.
\end{equation}

\section{Derivation of the optimal measurement operator}
\label{app:optimal}

In the context of a generalized quantum phase measurement where a positive-operator-valued measurement (POVM) $\{\hat{E}_{\xi}\}$ is carried out on the evolved probe state $\hat{\rho}_{\phi}$, the mean-square error of an unbiased estimator after $M$ measurement iterations is constrained from below by certain quantities
\begin{equation}
\Delta^2\phi 
\geq \frac{1}{M F(\phi)}
\geq \frac{1}{M H(\phi)}\,,
\label{eq:CRt}
\end{equation} 
where 
$F(\phi)=\int d\xi \frac{1}{p(\xi|\phi)}
(\frac{\partial p(\xi|\phi)}{\partial\phi})^2$ is the Fisher information  with $p(\xi|\phi)={\rm Tr}[\hat{\rho}_{\phi}\hat{E}_{\xi}]$ being the measurement probability associated with the outcome "$\xi$" 
and $H(\phi)={\rm Tr}[\hat{\rho}_{\phi}\hat{L}_{\phi}^2]$ is the quantum Fisher information with $\hat{L}_{\phi}$ being the symmetric logarithmic derivative operator defined as the Hermitian operator satisfying the following equation
\begin{equation}
\frac{\hat{L}_{\phi}\hat{\rho}_{\phi}+\hat{\rho}_{\phi}\hat{L}_{\phi}}{2}
=\frac{\partial \hat{\rho}_{\phi}}{\partial \phi}\,.
\label{eq:sld}
\end{equation}
The optimal estimator is the one that saturates the chain of inequalities in Eq.\,\eqref{eq:CRt}, and the conditions for achieving equality are provided below
\begin{equation}
\hat{E}_{\xi}\hat{\rho}_{\phi}=\lambda_{\xi,\phi}\hat{E}_{\xi}\hat{L}_{\phi}\hat{\rho}_{\phi}\,,
\end{equation}
in  which $\lambda_{\xi,\phi}={{\rm Tr}[\hat{\rho}_{\phi}\hat{E}_{\xi}]}/{{\rm Tr}[\hat{\rho}_{\phi}\hat{E}_{\xi}\hat{L}_{\phi}]}$ is a real number.
The fulfillment of the operatorial conditions outlined above can be ensured by constructing $\hat{E}_{\xi}$ from the set of projectors over the eigenstates of $\hat{L}_{\phi}$, where the generalized POVM reduces to the local projective von-Neuman measurements\,\cite{barndorff2000fisher,monras2006optimal,paris2009quantum}.
Specifically, the optimal estimator for determining a particular phase $\phi_t$ is $\hat{L}_{\phi}|_{\phi=\phi_t}$, which, using the error propagation formula, can be expressed as
\cite{hofmann2009achieving}
\begin{equation}
\begin{aligned}
\Delta^2\phi_{\hat{L}_\phi}|_{\phi=\phi_t}
& =\frac{\langle \hat{L}^2_\phi \rangle_{\phi_t}-\langle \hat{L}_\phi \rangle_{\phi_t}^2} 
{M |\partial \langle \hat{L}_\phi \rangle_{\phi_t}/\partial\phi|^2}|_{\phi=\phi_t}\\
& =\Delta^2\phi_{QCRB}\,,
\end{aligned}
\end{equation}
where $\langle \cdots \rangle_{\phi_t}$ is the expectation value for the evolved probe state in the absence of adaptive protocols.

\section{Theoretical verification of the validity of the optimal measurement scheme}
\label{app:veri}

In this part, we provide a verification for Eq.\,\eqref{eq:Ophi2} based on the error propagation formula.
The expectation value of $\hat{L}_0$ for the evolved probe state is given by
\begin{equation}
\langle \hat{L}_0 \rangle_{\delta\phi}=
\langle \hat{U}^{\dagger}_{\rm OPA_1}e^{i\delta\phi\hat{G}}\hat{U}_{\rm OPA_1}
\hat{L}_0
\hat{U}^{\dagger}_{\rm OPA_1}e^{-i\delta\phi\hat{G}}\hat{U}_{\rm OPA_1}\rangle_0\,,
\label{eq:L1}
\end{equation}
where $\langle \cdots \rangle_0$ represents the expectation value for the probe state.
Similarly, the expectation value of $\hat{L}^2_0$ for the evolved probe state is expressed as
\begin{equation}
\begin{aligned}
\langle \hat{L}^2_0 \rangle_{\delta\phi}
=\langle \hat{U}^{\dagger}_{\rm OPA_1}e^{i\delta\phi\hat{G}}\hat{U}_{\rm OPA_1}\hat{L}^{2}_0\hat{U}^{\dagger}_{\rm OPA_1}e^{-i\delta\phi\hat{G}}\hat{U}_{\rm OPA_1}\rangle_0\,.
\label{eq:L2}
\end{aligned}
\end{equation}
Furthermore, we establish that
\begin{align}
\frac{\partial \langle\hat{L}_0\rangle_{\delta\phi}}{\partial\phi}\nonumber
=
&\frac{1}{2}
[\langle \hat{U}^{\dagger}_{\rm OPA_1}e^{i\delta\phi\hat{G}}\hat{U}_{\rm OPA_1}\hat{L}_0\hat{U}^{\dagger}_{\rm OPA_1}e^{-i\delta\phi\hat{G}}\hat{U}_{\rm OPA_1}\hat{L}_0\rangle_0
\\
&+
\langle \hat{L}_0\hat{U}^{\dagger}_{\rm OPA_1}e^{i\delta\phi\hat{G}}\hat{U}_{\rm OPA_1}\hat{L}_0\hat{U}^{\dagger}_{\rm OPA_1}e^{-i\delta\phi\hat{G}}\hat{U}_{\rm OPA_1}\rangle_0
]\,,
\label{eq:DL1}
\end{align}
where Eq.\,\eqref{eq:sld} is utilized.
Given the smallness of $\delta\phi$, we can make an approximation as follows
\begin{equation}
e^{\pm i\delta\phi \hat{G}}\simeq 1 \pm i\delta\phi\hat{G}-\frac{\delta^2\phi}{2}\hat{G}^2\,.
\end{equation}
Subsequently, by leveraging the aforementioned approximation in Eq.\,\eqref{eq:L1}-Eq.\,\eqref{eq:DL1} and performing a series of mathematical operations on Eq.\,\eqref{eq:sld}, in our case, we obtain
\begin{equation}
\langle\hat{L}_0\rangle_{\delta\phi}
\simeq \delta\phi H(\phi)\,,
\end{equation}
\begin{equation}
\langle\hat{L}^2_0\rangle_{\delta\phi}
\simeq H(\phi)\,,
\end{equation}
\begin{equation}
\frac{\partial \langle\hat{L}_0\rangle_{\delta\phi}}{\partial\phi}
\simeq H(\phi)-\delta^2\phi H(\phi)
\frac{1}{2}(3\langle\hat{n}\rangle^2+6\langle\hat{n}\rangle+1)\,.
\end{equation}
Finally, the estimation variance of the optimal measurement scheme is determined to be
\begin{equation}
\Delta^2\phi_{\hat{L}_0}|_{\delta\phi\rightarrow 0}
\simeq
\frac{1}{MH(\phi)}
[1+\delta^2\phi(2\langle\hat{n}\rangle^2+4\langle\hat{n}\rangle+1)]\,.
\end{equation}
\nocite{*}
\bibliography{aipsamp}
\end{document}